\documentclass[11pt,a4paper]{article}
\usepackage[utf8]{inputenc}
\usepackage[T1]{fontenc}
\usepackage[english]{babel}
\usepackage[a4paper, left=1.25in, right=1.25in, top=1.35in, bottom=1.35in]{geometry}
\usepackage{graphicx}
\usepackage{amsmath,amssymb,amsthm,mathtools}
\usepackage{physics}
\usepackage{enumitem}
\usepackage{hyperref}
\usepackage{bm}
\usepackage{authblk}

\hypersetup{colorlinks=true, linkcolor=blue, citecolor=blue, urlcolor=blue}

\newcommand{\E}{\mathbb{E}}
\newcommand{\C}{\mathbb{C}}

\newcommand{\Ha}{\mathcal{H}}
\newcommand{\cG}{\mathcal{G}}

\newcommand{\Htwo}{\mathcal{H}^{(2)}}
\newcommand{\Rtwo}{\mathcal{R}^{(2)}}
\newcommand{\Gtwo}{\mathcal{G}^{(2)}}
\newcommand{\1}{\mathbf{1}}

\newtheorem{result}{Result}[section]
\newtheorem{definition}[result]{Definition}
\theoremstyle{remark}
\newtheorem{remark}[result]{Remark}

\title{A rank-two spherical transform\\ for eigenvector and singular-vector overlaps}
\author[1,2]{Pierre Bousseyroux\thanks{Email: pierre.bousseyroux@polytechnique.edu}}

\affil[1]{Econophysics Lab, Institut Louis Bachelier, 28 Pl. de la Bourse, Palais Brongniart, 75002 Paris, France}
\affil[2]{LadHyX, UMR CNRS 7646, Ecole Polytechnique, Institut Polytechnique de Paris, 91128 Palaiseau, France}
\date{}

\begin{document}
\maketitle

\begin{abstract}
We define a rank-two extension $\Htwo$ of the spherical
$\mathcal H$-transform introduced in \cite{bousseyroux2026r},
together with its associated transform $\Rtwo$.  For sums, both
transforms are additive.
The transform $\Rtwo$ yields a $4\times4$ formula for the average
product of two hermitized resolvents, and hence for off-diagonal
eigenvector overlaps, recovering the Ginibre, elliptic, and bi-invariant
formulas.  We also treat correlated pairs of matrices, and obtain overlaps
between their eigenvectors or singular vectors.
\end{abstract}

\section{Introduction}

Let $\mathbf M$ be an $N\times N$ diagonalizable non-Hermitian matrix with
simple eigenvalues $\lambda_1,\ldots,\lambda_N$.  We choose non-zero right and
left eigenvectors $\mathbf R_i,\mathbf L_i\in\C^N$ such that
\begin{equation}\label{eq:eigenvectors_intro}
    \mathbf M\mathbf R_i=\lambda_i\mathbf R_i,
    \qquad
    \mathbf L_i^*\mathbf M=\lambda_i\mathbf L_i^*,
    \qquad
    \mathbf L_i^*\mathbf R_j=\delta_{ij}.
\end{equation}
Here and below, $^*$ denotes conjugate transpose.
The Chalker--Mehlig overlaps are
\begin{equation}\label{eq:Oij}
    \mathcal O_{ij}
    :=
    (\mathbf L_i^*\mathbf L_j)(\mathbf R_j^*\mathbf R_i).
\end{equation}
They are invariant under the residual rescalings
$\mathbf R_i\mapsto c_i\mathbf R_i$ and
$\mathbf L_i\mapsto\bar c_i^{-1}\mathbf L_i$.  The diagonal terms are the
squared eigenvalue condition numbers, whereas the off-diagonal terms measure
correlations between two distinct biorthogonal eigendirections.  They govern
eigenvalue sensitivity and appear in the dynamics of non-Hermitian spectra
\cite{chalker1998eigenvector,mehlig2000statistical}.

Chalker and Mehlig first computed their one- and two-point densities for the
complex Ginibre ensemble
\cite{ginibre1965statistical,chalker1998eigenvector,mehlig2000statistical}.
Subsequent work obtained
precise distributions and microscopic asymptotics for Ginibre matrices
\cite{bourgade2020distribution} and related formulas for spherical and
truncated unitary ensembles \cite{dubach2021}.  Diagrammatic and
hermitization methods were
developed in parallel in the physics literature
\cite{janik1999correlations,janik1997non,nowak_tarnowski_2018}.  Explicit
macroscopic formulas remain concentrated in a few symmetric
families: Ginibre, elliptic, and bi-invariant ensembles.  Throughout this
article, rotational invariance means invariance in law under
$\mathbf M\mapsto\mathbf U\mathbf M\mathbf U^*$ for every deterministic
unitary matrix $\mathbf U$.
Here
\emph{bi-invariant} means invariance in law under
$\mathbf M\mapsto\mathbf U\mathbf M\mathbf V$ for every pair of deterministic
unitary matrices $\mathbf U,\mathbf V$.

Our starting point is the $\mathcal H$-transform introduced in
\cite{bousseyroux2026r}; its definition is recalled in
Section~\ref{sec:onepoint}.  The present paper applies the same spherical
integral construction to products of two resolvents.  We introduce four vectors
and define a rank-two transform $\Htwo_{\mathbf M}$.  Because the average is
taken after conjugation by a Haar unitary, $\Htwo_{\mathbf M}$ depends on
these vectors only through their Gram matrix.  The mixed second derivatives of
$\Htwo$ with respect to the cross-Gram block define a $4\times4$ transform
$\Rtwo_{\mathbf M}$.  Together with the one-point Green functions, this
transform determines the average of a product of two hermitized
resolvents by a single $4\times4$ inversion.  The definitions and the
two-point equation are given in
Sections~\ref{sec:onepoint}--\ref{sec:master}, with the replica
derivation in Appendix~\ref{app:proof}.

The same resolvent product also determines overlaps between singular vectors of
two correlated matrices.  The corresponding Hermitian formulas are
\cite{AllezBouchaud2014FreeAddition,BunBouchaudPotters2018Overlaps,potters2020first}.
We return to this after \eqref{eq:G2_master}.

\section{Setting and observables}\label{sec:setup}

Throughout the paper, the eigenvalues and biorthogonal eigenvectors of
$\mathbf M$ are those defined in \eqref{eq:eigenvectors_intro}, and
$\mathcal O_{ij}$ denotes \eqref{eq:Oij}. 

For a bulk point $z\in\C$, define the spectral density
\begin{equation}
    \rho_N(z):=\frac1N\E\sum_i \delta^{(2)}(z-\lambda_i),
    \qquad
    \rho(z):=\lim_{N\to\infty}\rho_N(z),
\end{equation}
and the normalized diagonal overlap density
\begin{equation}
    O_N^{(1)}(z):=\frac1{N^2}\E\sum_i \mathcal O_{ii}\,\delta^{(2)}(z-\lambda_i),
    \qquad
    O^{(1)}(z):=\lim_{N\to\infty}O_N^{(1)}(z).
\end{equation}
Here $\delta^{(2)}$ is the two-dimensional Dirac measure.
For the complex Ginibre ensemble with variance $1/N$, one has $\rho(z)=\pi^{-1}\mathbf 1_{|z|<1}$ and $O^{(1)}(z)=\pi^{-1}(1-|z|^2)\mathbf 1_{|z|<1}$, so that $\E[\mathcal O_{ii}\mid \lambda_i=z]\sim N(1-|z|^2)$ in the bulk: diagonal overlaps are of order $N$~\cite{chalker1998eigenvector,bourgade2020distribution}.

For two distinct points, we use the Chalker--Mehlig normalization
\begin{equation}
    O_N^{(2)}(z_1,z_2)
    :=
    \frac1N\E\sum_{i\neq j}\mathcal O_{ij}\,
    \delta^{(2)}(z_1-\lambda_i)\delta^{(2)}(z_2-\lambda_j),
    \qquad
    O^{(2)}(z_1,z_2)
    :=
    \lim_{N\to\infty}O_N^{(2)}(z_1,z_2).
\end{equation}
The
normalization of the one-point function above differs by one power of $N$
because it was chosen to have a finite bulk limit.  In the Ginibre bulk,
Chalker and Mehlig found
\begin{equation}\label{eq:CM}
    O^{(2)}_{\mathrm{Gin}}(z_1,z_2)
    =
    -\frac{1}{\pi^2}\,
    \frac{1-z_1\bar z_2}{|z_1-z_2|^4}
    \qquad z_1\neq z_2,
\end{equation}
inside the spectral domain~\cite{chalker1998eigenvector,mehlig2000statistical}.

\subsection[Hermitization and the 2 by 2 Green function]
{Hermitization and the $2\times2$ Green function}

Following Girko's hermitization method
\cite{girko1986elliptic,feinberg1997non,janik1997non}, introduce for
$\omega>0$ the $2N\times2N$ block matrix
\begin{equation}\label{eq:Cblock}
    \mathbf C(\omega,z)
    =
    \begin{pmatrix}
        \omega\1 & z\1-\mathbf M\\
        (z\1-\mathbf M)^* & \omega\1
    \end{pmatrix},
\end{equation}
and write its inverse in $N\times N$ blocks as
\begin{equation}
    \mathbf C(\omega,z)^{-1}=
    \begin{pmatrix}
        \mathbf G_{11} & \mathbf G_{12}\\
        \mathbf G_{21} & \mathbf G_{22}
    \end{pmatrix},
\end{equation}
the block traces
\begin{equation}
    \mathfrak g_{ij}^N(\omega,z):=\frac1N\Tr\mathbf G_{ij}(\omega,z)
\end{equation}
assemble into the $2\times2$ hermitized Green function
\cite{janik1997non,nowak_tarnowski_2018}
\begin{equation}
    \cG_{\mathbf M}^N(\omega,z)
    =
    \begin{pmatrix}
        \mathfrak g_{11}^N & \mathfrak g_{12}^N\\
        \mathfrak g_{21}^N & \mathfrak g_{22}^N
    \end{pmatrix},
    \qquad
    \cG_{\mathbf M}=\lim_{N\to\infty}\cG_{\mathbf M}^N.
\end{equation}
Write $\mathbf A=z\1-\mathbf M$.  The two diagonal blocks of
$\mathbf C(\omega,z)^{-1}$ are
$\omega(\omega^2\1-\mathbf A\mathbf A^*)^{-1}$ and
$\omega(\omega^2\1-\mathbf A^*\mathbf A)^{-1}$.  For a square matrix,
$\mathbf A\mathbf A^*$ and $\mathbf A^*\mathbf A$ have the same
eigenvalues, so
$\mathfrak g_{11}^N(\omega,z)=\mathfrak g_{22}^N(\omega,z)$.

Throughout the rank-two sector, compound indices are ordered as
\[
(21),\ (22),\ (11),\ (12).
\]

For two fixed points $(\omega_1,z_1)$ and $(\omega_2,z_2)$, define
\begin{equation}\label{eq:G2def}
\begin{aligned}
 \left[\Gtwo_{\mathbf M,N}\right]_{(ab),(cd)}
 &:=
 \frac1N\E\,\Tr\!\left[
 \mathbf G_{ca}(\omega_1,z_1)\,
 \mathbf G_{bd}(\omega_2,z_2)
 \right],
 \\
 \Gtwo_{\mathbf M}
 &:=
 \lim_{N\to\infty}\Gtwo_{\mathbf M,N}.
\end{aligned}
\end{equation}

In the eigenvalue problem one takes $\omega=-i\eta$ and lets
$\eta\downarrow0$.  The off-diagonal block then defines the Stieltjes
transform
\begin{equation}\label{eq:stieltjes}
 \mathfrak g(z)
 :=
 \lim_{\eta\downarrow0}\lim_{N\to\infty}
 \mathfrak g_{21}^N(-i\eta,z)
 =
 \int_{\C}\frac{\rho(w)}{z-w}\,\mathrm d^2w ,
\end{equation}
while
$\lim_{\eta\downarrow0}\lim_{N\to\infty}
\mathfrak g_{11}^N(-i\eta,z)=i\mathfrak o(z)$, with
$\mathfrak o(z)\ge0$ in the bulk.  The one-point resolvent
therefore determines both the spectral density and the diagonal overlap
density,
\begin{equation}\label{eq:diagoverlap}
    O^{(1)}(z)=\frac{\mathfrak o(z)^2}{\pi}
    \qquad\text{(bulk)},
\end{equation}
so that $\E[\mathcal O_{ii}\mid\lambda_i=z]
\sim N\mathfrak o(z)^2$
\cite{janik1999correlations,bousseyroux2026r}.

\subsection{Two-resolvent representation}

Define the \emph{two-point resolvent correlator}
\begin{equation}\label{eq:D}
    D_N(z_1,z_2)
    :=
    \frac1N\E\,\Tr\!\bigl[\mathbf G_{\mathbf M}(z_1)\,\mathbf G_{\mathbf M}(z_2)^*\bigr],
    \qquad
    D(z_1,z_2):=\lim_{N\to\infty}D_N(z_1,z_2),
\end{equation}
where $\mathbf G_{\mathbf M}(z):=(z\1-\mathbf M)^{-1}$.  Using the
spectral decomposition and biorthogonality,
\begin{equation}
    \Tr\!\bigl[\mathbf G_{\mathbf M}(z_1)
    \mathbf G_{\mathbf M}(z_2)^*\bigr]
    =
    \sum_{i,j}\frac{\mathcal O_{ij}}{(z_1-\lambda_i)(\bar z_2-\bar\lambda_j)}.
\end{equation}
By \eqref{eq:G2def}, $D$ is the $(11),(22)$ entry of
$\Gtwo_{\mathbf M}$.  The two-point equation for this entry is
\eqref{eq:G2_master}.
For $z_1\neq z_2$, the off-diagonal overlap is obtained by differentiating
$D$ with respect to $\bar z_1$ and $z_2$:
\begin{equation}\label{eq:OfromD}
    O^{(2)}(z_1,z_2)
    =
    \frac{1}{\pi^2}\,
    \partial_{\bar z_1}\partial_{z_2}\,
    D(z_1,z_2)
    \quad\text{(bulk)}.
\end{equation}
This follows from the distributional identity
\begin{equation}
 \partial_{\bar z}\frac1z=\pi\delta^{(2)}(z).
\end{equation}
For Ginibre, inserting the explicit formula
\begin{equation}\label{eq:DGin}
    D_{\mathrm{Gin}}(z_1,z_2)
    =
    \frac{1-|z_1|^2-|z_2|^2+\bar z_1z_2}{|z_1-z_2|^2},
    \qquad |z_a|<1,
\end{equation}
and applying~\eqref{eq:OfromD} reproduces the Chalker--Mehlig formula
\eqref{eq:CM}~\cite{chalker1998eigenvector,mehlig2000statistical}.  We
recover \eqref{eq:DGin} from the rank-two $\Htwo$-transform developed in
Sections~\ref{sec:rank2}--\ref{sec:master}.

\section[One-point recap: the H-transform]
{One-point recap: the $\Ha$-transform}\label{sec:onepoint}

\begin{definition}[$\mathcal H$-transform
  \cite{bousseyroux2026r}]
\label{def:H}
Let $\mathbf M$ be a deterministic or random $N\times N$ matrix and let
$\psi_1,\psi_2\in\C^N$.  The functional $\Ha_{\mathbf M}^N$ is defined by
\begin{equation}\label{eq:H_def}
\frac{1}{2N}\,
\log
\E_{\mathbf M}\E_{\mathbf U}\!\left[
\exp\!\Bigl(2N\,\Re\langle \psi_1,\,
\mathbf U\mathbf M\mathbf U^*\psi_2\rangle\Bigr)
\right]
=
\Ha_{\mathbf M}^N\bigl(
\|\psi_1\|\,\|\psi_2\|,\,
\langle\psi_1,\psi_2\rangle
\bigr),
\end{equation}
where $\mathbf U$ is Haar distributed on the unitary group
$\mathrm U(N)$ of $N\times N$ unitary matrices.  We write
$\Ha_{\mathbf M}:=\lim_{N\to\infty}\Ha_{\mathbf M}^N$ whenever this limit
exists.  Setting
\[
  \alpha=\|\psi_1\|\,\|\psi_2\|,
  \qquad
  \beta=\langle\psi_1,\psi_2\rangle,
\]
the associated transforms are
\begin{equation}\label{eq:R_transforms_def}
\mathcal R_1(\alpha,\beta)
:=\partial_\alpha\Ha(\alpha,\beta),
\qquad
\mathcal R_2(\alpha,\beta)
:=2\partial_\beta\Ha(\alpha,\beta),
\end{equation}
where
$\partial_\beta=\frac12(\partial_{\Re\beta}
-i\partial_{\Im\beta})$ is the Wirtinger derivative.
These derivatives are first obtained from arguments
$(\alpha,\beta)$ coming from a pair of vectors.  When they appear in the
Green-function equations below they may refer to another branch, reached
by an analytic continuation.
\end{definition}

Introduce the matrices
\begin{equation}\label{eq:matrix_ZGR}
 \mathbf Z(\omega,z)
 :=
 \begin{pmatrix}\omega&z\\ \bar z&\omega\end{pmatrix},
 \qquad
 \cG
 :=
 \begin{pmatrix}\mathfrak g_1&\overline{\mathfrak g_2}\\
 \mathfrak g_2&\mathfrak g_1\end{pmatrix},
 \qquad
 \mathbf R(\cG)
 :=
 \begin{pmatrix}
 \mathcal R_1&\mathcal R_2\\
 \overline{\mathcal R_2}&\mathcal R_1
 \end{pmatrix},
\end{equation}
where the transforms in the last matrix are evaluated at
$\alpha=\mathfrak g_1$ and $\beta=\mathfrak g_2$.  The placement of
$\mathcal R_2$ follows from the identification of the arguments of
$\Ha$ with the entries of the Gram matrix: at the one-point solution
$\beta=\langle\psi_1,\psi_2\rangle$ equals $\mathfrak g_{21}=\mathfrak g_2$,
while the $(1,2)$ entry of $\cG$ is $\overline{\mathfrak g_2}$; see
Appendix~\ref{app:proof}.  The one-point saddle-point equation of
\cite{bousseyroux2026r} can be written as
\begin{equation}\label{eq:one_point_relation}
 \cG(\omega,z)
 =
 \bigl[\mathbf Z(\omega,z)-\mathbf R(\cG(\omega,z))\bigr]^{-1}.
\end{equation}
Equation~\eqref{eq:one_point_relation} is used below in the two-point
equation.

For a sum $\mathbf A+\mathbf U\mathbf B\mathbf U^*$, independence and the
Haar integration give additivity of $\Ha$:
\begin{equation}
 \Ha_{\mathbf A+\mathbf U\mathbf B\mathbf U^*}
 =
 \Ha_{\mathbf A}+\Ha_{\mathbf B}.
\end{equation}
Equivalently, the one-point Green function obeys the subordination relation
\begin{equation}\label{eq:subord}
 \E_{\mathbf B}\E_{\mathbf U}\cG^N_{\mathbf A+\mathbf U\mathbf B\mathbf U^*}(\omega,z)
 =
 \cG_{\mathbf A}\!\left(
 \omega-\mathcal R_{1,\mathbf B}(\mathfrak g_1,\mathfrak g_2),
 z-\mathcal R_{2,\mathbf B}(\mathfrak g_1,\mathfrak g_2)
 \right).
\end{equation}
Here $\mathcal R_{1,\mathbf B}$ and $\mathcal R_{2,\mathbf B}$ may refer
to a particular branch.

\section[The rank-two H-transform]
{The rank-two $\Htwo$-transform}\label{sec:rank2}

\subsection{Two bilinear insertions and Gram matrices}

Consider four vectors
$\psi_1^1,\psi_2^1,\psi_1^2,\psi_2^2\in\C^N$, grouped into two bilinear
insertions
\begin{equation}
    \Psi^1=(\psi_1^1,\psi_2^1),
    \qquad
    \Psi^2=(\psi_1^2,\psi_2^2).
\end{equation}
We label a vector by the pair $(a,i)$, where $a\in\{1,2\}$ indexes the
bilinear insertion and $i\in\{1,2\}$ its position inside that insertion.
Their full Gram matrix is
\begin{equation}
 \mathbb G_{(a,i),(b,j)}
 :=
 \langle\psi_i^a,\psi_j^b\rangle,
 \qquad
 \mathbb G
 =
 \begin{pmatrix}G^1&C\\ C^*&G^2\end{pmatrix},
 \label{eq:full_gram}
\end{equation}
where $G^a=\Psi^{a*}\Psi^a$ and
$C=\Psi^{1*}\Psi^2$.

\begin{definition}[Rank-two $\mathcal H$-transform]\label{def:H2}
The rank-two transform $\Htwo_{\mathbf M,N}$ is defined by
\begin{equation}
 \frac1{2N}\log\E_{\mathbf U}\E_{\mathbf M}
 \exp\!\left\{
 2N\Re\sum_{a=1}^2
 \langle\psi_1^a,\mathbf U\mathbf M\mathbf U^*\psi_2^a\rangle
 \right\}
 =
 \Htwo_{\mathbf M,N}(\mathbb G).
 \label{eq:H2_def}
\end{equation}
Whenever the limit exists, we set
$\Htwo_{\mathbf M}=\lim_{N\to\infty}\Htwo_{\mathbf M,N}$.
\end{definition}

The average is taken after conjugation by a Haar unitary $\mathbf U$.  This
is what reduces the dependence on the four vectors to the Gram matrix
\eqref{eq:full_gram}, as
$(\psi_1,\psi_2)\mapsto(\alpha,\beta)$ in Definition~\ref{def:H}.  At the
block-diagonal configuration $\mathbb G_0=\operatorname{diag}(G^1,G^2)$,
the two insertions decouple:
\begin{equation}\label{eq:H2_decoupled}
 \Htwo_{\mathbf M}(\mathbb G_0)
 =
 \Ha_{\mathbf M}(G^1)+\Ha_{\mathbf M}(G^2),
\end{equation}
where $\Ha(G^a)$ is shorthand for
$\Ha(\sqrt{G^a_{11}G^a_{22}},G^a_{12})$.

\begin{definition}[Rank-two $\mathcal R$-transform]\label{def:R2}
Keep the diagonal Gram blocks fixed and differentiate with respect to the
cross-Gram matrix $C$.  Collect the independent entries as
\[
 \bm c
 :=
 \begin{pmatrix}
 C_{21}\\
 C_{22}\\
 C_{11}\\
 C_{12}
 \end{pmatrix}.
\]
Rows of $\Rtwo$ are indexed by $\overline{C_{ab}}$ and columns by
$C_{cd}$, in the order $(21),(22),(11),(12)$.  For $A,B\in M_2(\C)$,
\begin{equation}\label{eq:M2def}
 \left[
 \Rtwo_{\mathbf M}(A,B)
 \right]_{(ab),(cd)}
 :=
 \left.
 2\,
 \partial_{\overline{C_{ab}}}
 \partial_{C_{cd}}
 \Htwo_{\mathbf M}(\mathbb G)
 \right|_{
 C=0,\,
 G^1_{ij}=A_{ji},\,
 G^2_{ij}=B_{ji}
 }.
\end{equation}
Around $C=0$ one therefore has the expansion
\begin{equation}\label{eq:Fexpansion}
 \Htwo_{\mathbf M}(\mathbb G)
 =
 \Ha_{\mathbf M}(G^1)+\Ha_{\mathbf M}(G^2)
 +\frac12
 \bm c^*
 \Rtwo_{\mathbf M}(A,B)
 \bm c
 +o(\|C\|^2),
\end{equation}
where $A_{ij}=G^1_{ji}$ and $B_{ij}=G^2_{ji}$.
As for $\mathcal R_1$ and $\mathcal R_2$, we need to consider all
possible determinations.
\end{definition}

\begin{remark}
In the present non-Hermitian setting, the use of analytic
continuation of the $\mathcal R$-transforms, together with their
different determinations, was introduced in
\cite{bousseyroux2026r}.  This point of view provides a convenient
way of expressing spectral results directly in terms of suitable
branches of these transforms.
See
\cite{bousseyroux2026boundaries,bousseyroux2026outliers,bousseyroux2026multiplicative,bousseyroux2026multiplicativeBBP}.
\end{remark}

\subsection{Additivity}

For $\mathbf M=\mathbf A+\mathbf U\mathbf B\mathbf U^*$, with $\mathbf U$
Haar and independent of $\mathbf A,\mathbf B$, the two spherical factors
multiply, so
\begin{equation}
    \Htwo_{\mathbf M}=\Htwo_{\mathbf A}+\Htwo_{\mathbf B},
\end{equation}
hence
\begin{equation}\label{eq:addM2}
    \Rtwo_{\mathbf M}(A,B)
    =
    \Rtwo_{\mathbf A}(A,B)
    +
    \Rtwo_{\mathbf B}(A,B).
\end{equation}

\subsection{Complex Ginibre}

For the complex Ginibre ensemble \cite{ginibre1965statistical} with $\E[|M_{ij}|^2]=1/N$ and independent entries,
\begin{equation}
    \E\bigl[\Tr(\mathbf T\mathbf M)\Tr(\mathbf T^*\mathbf M^*)\bigr]
    =
    \frac1N\Tr(\mathbf T\mathbf T^*),
\end{equation}
for deterministic $\mathbf T$.  A direct Gaussian average in
\eqref{eq:H2_def} gives
\begin{equation}
    \Htwo_{\mathrm{Gin}}
    =
    \frac12G^1_{11}G^1_{22}
    +\frac12G^2_{11}G^2_{22}
    +\Re(C_{11}\overline{C_{22}}).
\end{equation}
Therefore, in the ordering of Definition~\ref{def:R2},
\begin{equation}\label{eq:GammaGin}
    \Rtwo_{\mathrm{Gin}}
    =
    \begin{pmatrix}
        0&0&0&0\\
        0&0&1&0\\
        0&1&0&0\\
        0&0&0&0
    \end{pmatrix}.
\end{equation}

\subsection{Elliptic Ginibre}

For the complex elliptic ensemble of
\cite{girko1986elliptic,sommers1988spectrum} with parameter
$\tau\in[-1,1]$, the
non-zero second moments are
\begin{equation}
 \E[M_{ij}\overline{M_{kl}}]
 =\frac1N\delta_{ik}\delta_{jl},
 \qquad
 \E[M_{ij}M_{kl}]
 =\frac{\tau}{N}\delta_{il}\delta_{jk}.
\end{equation}
A Gaussian average gives
\cite{girko1986elliptic,bousseyroux2026r}
\begin{equation}
    \Htwo_{\mathrm{ell}}
    =
    \Ha_{\mathrm{ell}}(G^1)+\Ha_{\mathrm{ell}}(G^2)
    +\Re(C_{11}\overline{C_{22}})
    +\tau\,\Re(C_{12}\overline{C_{21}}),
\end{equation}
so that
\begin{equation}\label{eq:GammaEll}
    \Rtwo_{\mathrm{ell}}
    =
    \begin{pmatrix}
        0&0&0&\tau\\
        0&0&1&0\\
        0&1&0&0\\
        \tau&0&0&0
    \end{pmatrix}.
\end{equation}
Again $\Rtwo$ is constant.  

\subsection{Bi-invariant reduction via singular values}

If $\mathbf M$ is bi-invariant, the one-point theory collapses to a single
scalar function of one variable
\cite{bousseyroux2026r,benaych2011rectangular}: writing
$\mu_s$ for the limiting symmetrized singular-value law of $\mathbf M$,
\begin{equation}\label{eq:R_biinv_onepoint}
 \mathbf R(\cG)
 =
 \mathcal R_1(\mathfrak g_1)\,\1_2,
 \qquad
 \mathcal R_2\equiv0,
\end{equation}
where $\mathcal R_1=R_{\mu_s}$ is the free additive $R$-transform of $\mu_s$.
We use the fixed-rank extension of the rectangular spherical-integral
formula \cite{benaych2011rectangular}.  Let
$\mathbf T=\bm a_1\bm a_2^*+\bm a_3\bm a_4^*$, and let $s_1,s_2$ be
its singular values.
Then
\begin{equation}
    \Htwo_{\mathbf M}(\mathbb G)
    =
    H(s_1)+H(s_2),
\end{equation}
where $H'(x)=\mathcal R_1(x)$.
At the block-diagonal saddle,
$G^a_{11}=G^a_{22}=:x_a$.  Expanding $s_1,s_2$ at $C=0$ and matching
\eqref{eq:Fexpansion} yields
\begin{equation}\label{eq:R2_biinv_early}
    \Rtwo_{\mathrm{biinv}}
    =
    \begin{pmatrix}
        0 & 0 & 0 & 0\\
        0 & \kappa_d & \kappa_x & 0\\
        0 & \kappa_x & \kappa_d & 0\\
        0 & 0 & 0 & 0
    \end{pmatrix},
\end{equation}
with
\begin{equation}\label{eq:kappa_biinv}
    \kappa_d
    =
    \frac{x_2\,\mathcal R_1(x_1)-x_1\,\mathcal R_1(x_2)}{x_1^2-x_2^2},
    \qquad
    \kappa_x
    =
    \frac{x_1\,\mathcal R_1(x_1)-x_2\,\mathcal R_1(x_2)}{x_1^2-x_2^2},
\end{equation}
with the continuous values
\begin{equation}
 \kappa_d(x,x)=\frac{x\mathcal R_1'(x)-\mathcal R_1(x)}{2x},
 \qquad
 \kappa_x(x,x)=\frac{x\mathcal R_1'(x)+\mathcal R_1(x)}{2x}.
\end{equation}
At $x=0$, these expressions are again understood by continuity.  For Ginibre,
$\mathcal R_1(x)=x$, so $\kappa_d=0$ and $\kappa_x=1$, recovering
\eqref{eq:GammaGin}.

\section{Two-point equation}\label{sec:master}

\begin{result}[Two-point resolvent equation]\label{result:master}
Let $\mathbf M$ be a rotationally invariant
non-Hermitian matrix, and
let $\cG_a=\cG_{\mathbf M}(\omega_a,z_a)$ be its $2\times2$ hermitized Green
function at the points $(\omega_1,z_1)$ and $(\omega_2,z_2)$.  The
$4\times4$ matrix defined in
\eqref{eq:G2def} is
\begin{equation}
    \Gtwo_{\mathbf M}
    =
    \left[
    (\cG_1\otimes\cG_2)^{-1}
    -
    \Rtwo_{\mathbf M}(\cG_1,\cG_2)
    \right]^{-1}.
\label{eq:G2_master}
\end{equation}
Equivalently,
\begin{equation}\label{eq:G2_ZR}
 \Gtwo_{\mathbf M}
 =
 \left[
 \bigl(\mathbf Z_1-\mathbf R(\cG_1)\bigr)
 \otimes
 \bigl(\mathbf Z_2-\mathbf R(\cG_2)\bigr)
 -
 \Rtwo_{\mathbf M}(\cG_1,\cG_2)
 \right]^{-1}.
\end{equation}
Here and below, $\mathbf R$ and $\Rtwo_{\mathbf M}$ are understood on suitable branches of the corresponding transforms.
\end{result}

A replica derivation of \eqref{eq:G2_master} is given in
Appendix~\ref{app:proof}.

\begin{remark}
The quantity $D(z_1,z_2)$ in \eqref{eq:D} is the
$(11),(22)$ entry of $\Gtwo_{\mathbf M}$
in the limit $\omega_1,\omega_2\to0$.
The off-diagonal overlap then follows from \eqref{eq:OfromD}.
\end{remark}

Outside the spectrum, set $g_a:=\mathfrak g_{\mathbf M}(z_a)$.
At $\omega=0$,
\begin{equation}\label{eq:exterior_G}
 \cG_a
 =
 \begin{pmatrix}
 0&\overline{g_a}\\
 g_a&0
 \end{pmatrix},
 \qquad a=1,2.
\end{equation}
Hence
\begin{equation}\label{eq:exterior_bare}
 \cG_1\otimes\cG_2
 =
 \begin{pmatrix}
 0&0&0&\overline{g_1}\,\overline{g_2}\\
 0&0&\overline{g_1}g_2&0\\
 0&g_1\overline{g_2}&0&0\\
 g_1g_2&0&0&0
 \end{pmatrix}.
\end{equation}
The $(11),(22)$ entry of \eqref{eq:G2_master} is therefore
\begin{equation}\label{eq:exterior_inverse}
 \frac{1}{D(z_1,z_2)}
 =
 \frac{1}{\mathfrak g_{\mathbf M}(z_1)\overline{\mathfrak g_{\mathbf M}(z_2)}}
 -
 \left[
 \Rtwo_{\mathbf M}(\cG_1,\cG_2)
 \right]_{(22),(11)}.
\end{equation}

\begin{result}\label{result:exterior}
Let $\mathbf M$ be a rotationally invariant non-Hermitian matrix,
and let $z_1,z_2$ lie outside the spectrum.  Then
\begin{equation}\label{eq:exterior_R2}
\begin{aligned}
 &\Biggl[
 \Rtwo_{\mathbf M}\!
 \Biggl(
 \begin{pmatrix}
 0&\overline{\mathfrak g_{\mathbf M}(z_1)}\\
 \mathfrak g_{\mathbf M}(z_1)&0
 \end{pmatrix},
 \begin{pmatrix}
 0&\overline{\mathfrak g_{\mathbf M}(z_2)}\\
 \mathfrak g_{\mathbf M}(z_2)&0
 \end{pmatrix}
 \Biggr)
 \Biggr]_{(22),(11)}
 \\
 &\qquad
 =
 \frac{1}{\mathfrak g_{\mathbf M}(z_1)\overline{\mathfrak g_{\mathbf M}(z_2)}}
 -
 \frac{1}{D(z_1,z_2)}.
\end{aligned}
\end{equation}
Here and below, $\Rtwo_{\mathbf M}$ denotes a suitable determination of the rank-two transform.
\end{result}

\begin{remark}
For $z_1=z_2=z$, the left-hand side of \eqref{eq:exterior_R2} equals
$\partial_\alpha\mathcal R_{1,\mathbf M}(0,\mathfrak g_{\mathbf M}(z))$,
and hence
\begin{equation}\label{eq:exterior_R1}
 \partial_\alpha\mathcal R_{1,\mathbf M}
 \bigl(0,\mathfrak g_{\mathbf M}(z)\bigr)
 =
 \frac{1}{|\mathfrak g_{\mathbf M}(z)|^2}
 -
 \frac{1}{D(z,z)}.
\end{equation}
Compare \cite[Eq.~(36)]{bousseyroux2026r}.
\end{remark}

\begin{remark}
For $z$ in the same component of the resolvent set, the one-point
equation gives
\begin{equation}\label{eq:R_scalar_relation}
 \mathcal R_{2,\mathbf M}
 \bigl(0,\mathfrak g_{\mathbf M}(z)\bigr)
 =
 z-\frac{1}{\mathfrak g_{\mathbf M}(z)}.
\end{equation}
\end{remark}

\begin{remark}[Bi-invariant case]
For a bi-invariant matrix, \eqref{eq:R2_biinv_early} gives
\[
 \left[\Rtwo_{\mathrm{biinv}}\right]_{(22),(11)}
 =\kappa_x(0,0)
 =\mathcal R_1'(0)
 =\tau(\mathbf M\mathbf M^*),
\]
where
$\tau(\mathbf M\mathbf M^*)=\lim_{N\to\infty}\frac1N\E\Tr(\mathbf M\mathbf M^*)$.
The single ring theorem identifies this with the squared outer radius
of the eigenvalue support
\cite{feinberg1997non,haagerup2000brown,guionnet2011single}.
Since $\mathfrak g_{\mathbf M}(z)=1/z$ on the unbounded component of the resolvent set, i.e. outside the outer boundary of the eigenvalue support,
\eqref{eq:exterior_R2} yields
\begin{equation}\label{eq:D_biinv_exterior}
 D_{\mathrm{biinv}}(z_1,z_2)
 =
 \frac{1}{z_1\bar z_2-\tau(\mathbf M\mathbf M^*)}.
\end{equation}
This is \cite[Eq.~(3.32)]{nowak_tarnowski_2018}.
\end{remark}

\section{Examples}\label{sec:examples}

\subsection{Complex Ginibre}\label{sec:ginibre}

Inside the unit disk, the one-point solution of
\cite{bousseyroux2026r} is
\begin{equation}
    \mathfrak g(z)=\bar z,
    \qquad
    \mathfrak o(z)=\sqrt{1-|z|^2},
\end{equation}
so that
\begin{equation}
    \cG(z)
    =
    \begin{pmatrix}
        i\mathfrak o(z) & \overline{\mathfrak g(z)}\\
        \mathfrak g(z) & i\mathfrak o(z)
    \end{pmatrix}
    =
    \begin{pmatrix}
        i\mathfrak o(z) & z\\
        \bar z & i\mathfrak o(z)
    \end{pmatrix},
    \qquad
    \cG(z)^{-1}
    =
    \begin{pmatrix}
        -i\mathfrak o(z) & z\\
        \bar z & -i\mathfrak o(z)
    \end{pmatrix}.
\end{equation}
With $\Rtwo_{\mathrm{Gin}}$ as in \eqref{eq:GammaGin},
Equation~\eqref{eq:G2_master} gives
\begin{equation}\label{eq:Ginibre_explicit}
 D_{\mathrm{Gin}}(z_1,z_2)
 =
 \bigl[(\cG_1\otimes\cG_2)^{-1}-\Rtwo_{\mathrm{Gin}}\bigr]^{-1}_{(11),(22)}
 =
 \frac{1-|z_1|^2-|z_2|^2+\bar z_1z_2}{|z_1-z_2|^2},
\end{equation}
which is~\eqref{eq:DGin}.  Applying~\eqref{eq:OfromD} reproduces the
Chalker--Mehlig formula~\eqref{eq:CM}.

\subsection{Elliptic Ginibre}

For the elliptic ensemble
\cite{girko1986elliptic,sommers1988spectrum},
$\mathcal R_1(\alpha,\beta)=\alpha$ and
$\mathcal R_2(\alpha,\beta)=\tau\beta$
\cite{bousseyroux2026r}.  We restrict here to $|\tau|<1$; the endpoints
$\tau=\pm1$ correspond to degenerate one-dimensional limits and are not
described by the two-dimensional bulk formulas below.
Inside the spectral ellipse,
\begin{equation}
    \mathfrak g(z)=\frac{\bar z-\tau z}{1-\tau^2},
    \qquad
    \mathfrak o(z)=\sqrt{1-|\mathfrak g(z)|^2}.
\end{equation}
The transform $\Rtwo_{\mathrm{ell}}$ of \eqref{eq:GammaEll} is
constant.  Writing
\[
 \cG(z)
 =
 \begin{pmatrix}
 i\mathfrak o(z)&\overline{\mathfrak g(z)}\\
 \mathfrak g(z)&i\mathfrak o(z)
 \end{pmatrix}
\]
the $4\times4$ inversion in \eqref{eq:G2_master} can be performed
explicitly.  For two distinct bulk points and $|\tau|<1$, it gives
\begin{equation}\label{eq:Dell}
 D_{\mathrm{ell}}(z_1,z_2)
 =
 \frac{
 (1-\tau^2)^2
 -|z_1|^2-|z_2|^2
 +\bar z_1 z_2
 +\tau(z_1^2+\bar z_2^2)
 -\tau^2 z_1\bar z_2
 }{
 (1-\tau^2)|z_1-z_2|^2
 }.
\end{equation}
Consequently,
\begin{equation}\label{eq:Oell}
 O_{\mathrm{ell}}^{(2)}(z_1,z_2)
 =
 -\frac{
 (1-\tau^2)^2
 +\tau(z_1^2+\bar z_2^2)
 -(1+\tau^2)z_1\bar z_2
 }{
 \pi^2(1-\tau^2)|z_1-z_2|^4
 }.
\end{equation}
This recovers the known macroscopic two-point eigenvector-overlap
formula for the complex elliptic ensemble
\cite{mehlig2000statistical,nowak_tarnowski_2018}.
For $\tau=0$, it reduces immediately to the Chalker--Mehlig result
\eqref{eq:CM}.

\subsection{Bi-invariant case}\label{sec:biinv}

Let $\mathbf M$ be bi-invariant.  Its limiting eigenvalue density is
rotationally symmetric.  We write
\begin{equation}\label{eq:F_cdf}
 F(r):=2\pi\int_0^r \rho(s)\,s\,ds,
 \qquad r=|z|,
\end{equation}
so that $F(r)$ is the limiting fraction of eigenvalues in the disk of
radius $r$.  For a rotationally symmetric density, angular integration
gives, for $z\neq0$,
\begin{equation}
 \mathfrak g(z)=\frac{F(r)}{z}.
\end{equation}

In the bi-invariant case the one-point $R$-transform is scalar,
\[
 \mathbf R(\cG)=\mathcal R_1(\mathfrak g_1)\1_2,
 \qquad \mathcal R_2\equiv0.
\]
After the continuation $\omega=-i\eta$, $\eta\downarrow0$, write
\[
 x:=i\mathfrak o(z),
 \qquad
 \cG(z)=
 \begin{pmatrix}
 x&F(r)/\bar z\\
 F(r)/z&x
 \end{pmatrix}.
\]
The one-point equation
$\cG^{-1}=\mathbf Z-\mathbf R(\cG)$ therefore reads
\[
 \cG(z)^{-1}
 =
 \begin{pmatrix}
 -\mathcal R_1(x)&z\\
 \bar z&-\mathcal R_1(x)
 \end{pmatrix}.
\]
On the other hand,
\[
 \cG(z)^{-1}
 =
 \frac{1}{x^2-F(r)^2/r^2}
 \begin{pmatrix}
 x&-F(r)/\bar z\\
 -F(r)/z&x
 \end{pmatrix}.
\]
Comparison of the off-diagonal entries gives
\begin{equation}\label{eq:F_relation}
 x^2=-\frac{F(r)(1-F(r))}{r^2},
 \qquad
 \mathfrak o(z)^2
 =\frac{F(r)(1-F(r))}{r^2},
\end{equation}
while comparison of the diagonal entries gives the useful identity
\begin{equation}\label{eq:R1_F_relation}
 \mathcal R_1(x)
 =
 \frac{x r^2}{F(r)}
 =
 -\frac{1-F(r)}{x}.
\end{equation}
In particular, using \eqref{eq:diagoverlap},
\begin{equation}\label{eq:O1_bi}
 O^{(1)}(r)
 =
 \frac{F(r)(1-F(r))}{\pi r^2}.
\end{equation}
This is the formula of
\cite{belinschi2017squared}.

We now consider two bulk points $z_1,z_2$ and set
\[
 r_a=|z_a|,
 \qquad
 F_a=F(r_a),
 \qquad
 x_a=i\mathfrak o(z_a),
 \qquad
 u_a=\frac{F_a}{\bar z_a},
 \qquad
 v_a=\frac{F_a}{z_a}.
\]
The transform $\Rtwo$ derived in
\eqref{eq:R2_biinv_early} is
\[
 \Rtwo_{\mathrm{biinv}}
 =
 \begin{pmatrix}
 0&0&0&0\\
 0&\kappa_d&\kappa_x&0\\
 0&\kappa_x&\kappa_d&0\\
 0&0&0&0
 \end{pmatrix},
\]
where
\[
 \kappa_d
 =
 \frac{x_2\mathcal R_1(x_1)-x_1\mathcal R_1(x_2)}
 {x_1^2-x_2^2},
 \qquad
 \kappa_x
 =
 \frac{x_1\mathcal R_1(x_1)-x_2\mathcal R_1(x_2)}
 {x_1^2-x_2^2}.
\]
Using \eqref{eq:R1_F_relation},
\begin{equation}\label{eq:kappa_F}
 \kappa_x
 =
 \frac{F_1-F_2}{x_1^2-x_2^2},
 \qquad
 \kappa_d x_1x_2
 =
 \frac{x_1^2(1-F_2)-x_2^2(1-F_1)}
 {x_1^2-x_2^2}.
\end{equation}

Substituting this kernel into the two-point equation
\eqref{eq:G2_master}, the $(11),(22)$ component can be written as
\begin{equation}\label{eq:D_biinv_intermediate}
 D_{\mathrm{bi}}(z_1,z_2)
 =
 \frac{
 u_2v_1+
 \kappa_x
 (x_1^2x_2^2-u_1u_2v_1v_2)}
 {\Delta},
\end{equation}
with
\[
 \Delta
 =
 1-2\kappa_d x_1x_2
 -\kappa_x(u_1v_2+u_2v_1)
 +(\kappa_d^2-\kappa_x^2)
 (x_1^2x_2^2-u_1u_2v_1v_2).
\]
Using \eqref{eq:F_relation} and \eqref{eq:kappa_F}, this simplifies to
\begin{equation}\label{eq:D_biinv_closed}
 D_{\mathrm{bi}}(z_1,z_2)
 =
 \frac{
 F_1(1-F_1)(z_1-z_2)/z_1
 +
 F_2(1-F_2)(\bar z_1-\bar z_2)/\bar z_2}
 {|z_1-z_2|^2(F_1-F_2)}.
\end{equation}
Equivalently, by \eqref{eq:O1_bi},
\begin{equation}
 D_{\mathrm{bi}}(z_1,z_2)
 =
 \pi\,
 \frac{
 \bar z_1(z_1-z_2)O^{(1)}(r_1)
 +
 z_2(\bar z_1-\bar z_2)O^{(1)}(r_2)}
 {|z_1-z_2|^2\,\bigl[F(r_1)-F(r_2)\bigr]}.
\end{equation}

Finally, \eqref{eq:OfromD} gives
\begin{equation}\label{eq:O2_NT}
 O^{(2)}_{\mathrm{bi}}(z_1,z_2)
 =
 \frac1\pi
 \partial_{\bar z_1}\partial_{z_2}
 \left[
 \frac{
 \bar z_1(z_1-z_2)O^{(1)}(r_1)
 +
 z_2(\bar z_1-\bar z_2)O^{(1)}(r_2)}
 {|z_1-z_2|^2\,\bigl[F(r_1)-F(r_2)\bigr]}
 \right].
\end{equation}
This is the general bi-invariant formula of
\cite{nowak_tarnowski_2018}.

As a check, for complex Ginibre $F(r)=r^2$ in the unit disk, and
\eqref{eq:D_biinv_closed} reduces to
\eqref{eq:Ginibre_explicit}; consequently \eqref{eq:O2_NT} gives the
Chalker--Mehlig formula \eqref{eq:CM}.  Other bi-invariant examples,
such as induced Ginibre, truncated unitary matrices and products of
Ginibre matrices, follow by inserting their corresponding radial
cumulative distribution function $F$.

\subsection{Elliptic plus a product of two Ginibre matrices}
\label{sec:hybrid}

We now consider the additive example
\begin{equation}
 \mathbf M=\mathbf E_\tau+\mathbf B,
 \qquad
 \mathbf B=\mathbf Y_1\mathbf Y_2,
\end{equation}
where $\mathbf E_\tau$ is a complex elliptic matrix with parameter $\tau$,
$\mathbf Y_1$ and $\mathbf Y_2$ are independent complex Ginibre matrices, and
\begin{equation}
 \E|(\mathbf Y_a)_{ij}|^2=\frac{\sigma_a^2}{N},
 \qquad
 v=\sigma_1^2\sigma_2^2.
\end{equation}
These three matrices are independent.  The product $\mathbf B$ is
bi-invariant, and the free additive transform of its symmetrized singular
values is \cite{bousseyroux2026r}
\begin{equation}\label{eq:R_product}
 \mathcal R_{1,\mathbf B}(x)=\frac{vx}{1-vx^2}.
\end{equation}
Write the one-point solution in the bulk as
\[
 \cG(z)=
 \begin{pmatrix}h&\gamma\\ \bar\gamma&h\end{pmatrix},
 \qquad h=i\sqrt{s},
 \qquad \chi=s+|\gamma|^2.
\]
The bulk solution is the root $\chi$ of
\begin{equation}\label{eq:chi_hybrid}
 \frac{(1+v)\chi-1}{v(1-\chi)}
 +\chi^2\left[
 \frac{(\Re z)^2}{(1+\tau\chi)^2}
 +
 \frac{(\Im z)^2}{(1-\tau\chi)^2}
 \right]-\chi=0
\end{equation}
for which $s=((1+v)\chi-1)/(v(1-\chi))$ is non-negative.  Then
\begin{equation}\label{eq:gamma_hybrid}
 \gamma(z)
 =
 \frac{\chi z-\tau\chi^2\bar z}{1-\tau^2\chi^2}.
\end{equation}
The boundary is the ellipse
\begin{equation}
 \frac{(\Re z)^2}{(\sqrt{1+v}+\tau/\sqrt{1+v})^2}
 +
 \frac{(\Im z)^2}{(\sqrt{1+v}-\tau/\sqrt{1+v})^2}=1.
\end{equation}

\begin{figure}[t]
 \centering
 \includegraphics[width=\textwidth]{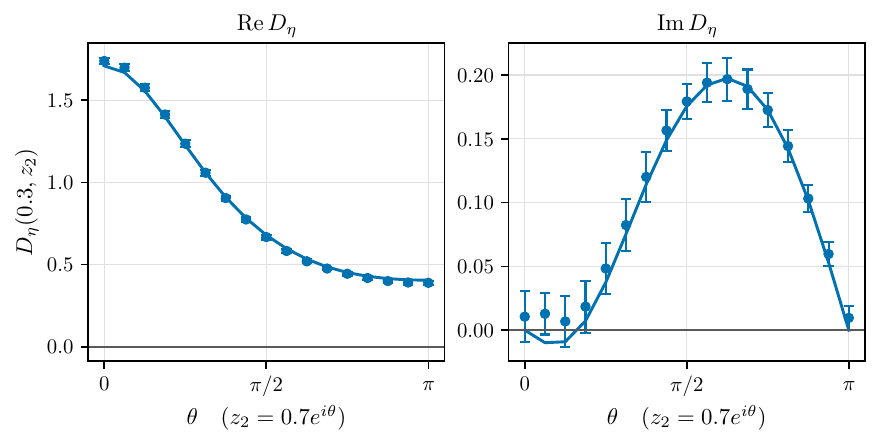}
 \caption{Real and imaginary parts of
 $D_\eta(z_1,z_2)
 =\frac1N\Tr\bigl[
 \mathbf G_{21}(-i\eta,z_1)\,
 \mathbf G_{12}(-i\eta,z_2)
 \bigr]$
 against $\theta$, for
 $\mathbf M=\mathbf E_{0.5}+\mathbf Y_1\mathbf Y_2$,
 $z_1=0.3$, $z_2=0.7e^{i\theta}$, $v=1$ and $\eta=0.12$.
 Line: $(11),(22)$ entry of \eqref{eq:G2_master} at $\omega=-i\eta$.
 Points: same trace averaged over $25$ matrices of size $N=60$.}
 \label{fig:hybrid-correlator}
\end{figure}

For two spectral points let $h_a=h(z_a)$,
$L_a=1-vh_a^2$, and set
\begin{equation}\label{eq:R2_hybrid}
 \Rtwo_{\mathbf M}(\cG_1,\cG_2)
 =
 \begin{pmatrix}
 0&0&0&\tau\\
 0&d&a&0\\
 0&a&d&0\\
 \tau&0&0&0
 \end{pmatrix},
 \qquad
 a=1+\frac{v}{L_1L_2},
 \qquad
 d=\frac{v^2h_1h_2}{L_1L_2}.
\end{equation}
Then
\begin{equation}\label{eq:O_hybrid}
 D_{\mathrm{hyb}}(z_1,z_2)
 =
 \left[
 (\cG_1\otimes\cG_2)^{-1}-\Rtwo_{\mathbf M}(\cG_1,\cG_2)
 \right]^{-1}_{(11),(22)},
 \qquad
 O_{\mathrm{hyb}}^{(2)}
 =
 \frac1{\pi^2}\partial_{\bar z_1}\partial_{z_2}D_{\mathrm{hyb}} .
\end{equation}
Evaluation requires solving \eqref{eq:chi_hybrid} at each spectral point
and then performing the $4\times4$ inversion in \eqref{eq:O_hybrid}.
Figure~\ref{fig:hybrid-correlator} compares the $(11),(22)$ entry of
\eqref{eq:G2_master} at $\omega_1=\omega_2=-i\eta$ with the finite-$N$
average of
$\frac1N\Tr\bigl[\mathbf G_{21}(-i\eta,z_1)\,\mathbf G_{12}(-i\eta,z_2)\bigr]$.
We fix $z_1=0.3$ and move the second point along the semicircle
$z_2=0.7e^{i\theta}$.

\section{Two matrices}\label{sec:twomatrix}

Let $(\mathbf X,\mathbf Y)$ have a joint law invariant under simultaneous
unitary conjugation,
\[
 (\mathbf X,\mathbf Y)
 \stackrel{\mathrm d}=
 (\mathbf U\mathbf X\mathbf U^*,\mathbf U\mathbf Y\mathbf U^*).
\]
We define the associated rank-two transform by
\begin{equation}\label{eq:H2_pair}
 \frac1{2N}\log\E
 \exp\!\left\{2N\Re\left(
 \langle\psi_1^1,\mathbf X\psi_2^1\rangle+
 \langle\psi_1^2,\mathbf Y\psi_2^2\rangle\right)\right\}
 =
 \Htwo_{\mathbf X,\mathbf Y;N}(\mathbb G).
\end{equation}
\begin{result}[Two-matrix formula]\label{result:mixed-pair}
Write \(\Rtwo_{\mathbf X,\mathbf Y}\) for the Hessian of
\eqref{eq:H2_pair}, defined as in Definition~\ref{def:R2}, and
$\cG_{\mathbf X}:=\cG_{\mathbf X}(\omega_1,z_1)$,
$\cG_{\mathbf Y}:=\cG_{\mathbf Y}(\omega_2,z_2)$.  Then
\begin{equation}\label{eq:G2_pair}
 \Gtwo_{\mathbf X,\mathbf Y}
 =
 \left[
 (\cG_{\mathbf X}\otimes\cG_{\mathbf Y})^{-1}
 -
 \Rtwo_{\mathbf X,\mathbf Y}
 (\cG_{\mathbf X},\cG_{\mathbf Y})
 \right]^{-1}.
\end{equation}
\end{result}
The replica computation of Appendix~\ref{app:proof} carries over with
$\mathbf M$ replaced by $\mathbf X$ and $\mathbf Y$ in the two bilinear
insertions.

\subsection[The pair (A, A+B)]
{The pair $(\mathbf A,\mathbf A+\mathbf B)$}

Let $\mathbf A$ and $\mathbf B$ be independent, and consider the pair
$(\mathbf A,\mathbf A+\mathbf U\mathbf B\mathbf U^*)$ with $\mathbf U$ Haar.
In the spherical integral, the contribution of $\mathbf B$ contains only
the second bilinear insertion.  Consequently,
\begin{equation}\label{eq:H2_A_AplusB}
 \Htwo_{\mathbf A,\,\mathbf A+\mathbf U\mathbf B\mathbf U^*}
 (\mathbb G)
 =
 \Htwo_{\mathbf A,\mathbf A}(\mathbb G)
 +
 \Ha_{\mathbf B}(G^2),
\end{equation}
and therefore
\begin{equation}\label{eq:R2_A_AplusB}
 \Rtwo_{\mathbf A,\,\mathbf A+\mathbf U\mathbf B\mathbf U^*}
 =
 \Rtwo_{\mathbf A,\mathbf A}.
\end{equation}
The added matrix still changes $\cG_{\mathbf A+\mathbf U\mathbf B\mathbf U^*}$
through the one-point subordination equation \eqref{eq:subord}, and therefore
changes the overlaps.

\subsection{Mixed biorthogonal overlaps}\label{sec:mixed-overlaps}

The quantity generated by a product of two \emph{ordinary} resolvents, one
for each matrix of the pair, is the mixed biorthogonal overlap
\begin{equation}\label{eq:Q_mixed}
 \mathcal Q_{ij}^{\mathbf X,\mathbf Y}
 :=
 \bigl((\mathbf L_i^{\mathbf X})^*\mathbf L_j^{\mathbf Y}\bigr)
 \bigl((\mathbf R_j^{\mathbf Y})^*\mathbf R_i^{\mathbf X}\bigr),
\end{equation}
where $\mathbf L_i^{\mathbf X},\mathbf R_i^{\mathbf X}$ are the left and
right eigenvectors of $\mathbf X$ associated with $\lambda_i$, normalized by
$(\mathbf L_i^{\mathbf X})^*\mathbf R_i^{\mathbf X}=1$, and similarly for
$\mathbf Y$ with eigenvalues $\mu_j$.  It is invariant under the allowed
eigenvector rescalings and reduces to $\mathcal O_{ij}$ when
$\mathbf X=\mathbf Y$.  Repeating the computation of \eqref{eq:D} for the
pair gives
\begin{equation}\label{eq:D_pair}
 D_{\mathbf X,\mathbf Y}(z_1,z_2)
 :=
 \frac1N\E\Tr\!\left[
 (z_1\1-\mathbf X)^{-1}
 \bigl((z_2\1-\mathbf Y)^{-1}\bigr)^*
 \right]
 =
 \frac1N\E\sum_{i,j}
 \frac{\mathcal Q_{ij}^{\mathbf X,\mathbf Y}}
 {(z_1-\lambda_i)(\bar z_2-\bar\mu_j)},
\end{equation}
which is the $(11),(22)$ entry of \eqref{eq:G2_pair}.  At
$\omega_1=\omega_2=-i\eta$ the same entry is built from the $(21)$ block
of $\mathbf X$ and the $(12)$ block of $\mathbf Y$.
\begin{equation}\label{eq:Q_conditional}
 N\,\E\!\left[
 \mathcal Q_{ij}^{\mathbf X,\mathbf Y}
 \,\middle|\,
 \lambda_i=z_1,\ \mu_j=z_2
 \right]
 =
 \frac{\partial_{\bar z_1}\partial_{z_2}D_{\mathbf X,\mathbf Y}(z_1,z_2)}
 {\pi^2\rho_{\mathbf X}(z_1)\rho_{\mathbf Y}(z_2)}.
\end{equation}

\subsubsection*{Two Ginibre matrices}

Take $\mathbf X=\mathbf A$ and $\mathbf Y=\mathbf A+\mathbf B$ with
$\mathbf A,\mathbf B$ independent complex Ginibre matrices,
$\E|A_{ij}|^2=a/N$, $\E|B_{ij}|^2=b/N$, and $c=a+b$.  By
\eqref{eq:R2_A_AplusB}, $\Rtwo_{\mathbf X,\mathbf Y}=a\,\Rtwo_{\mathrm{Gin}}$,
while the two one-point factors differ:
\[
 \cG_{\mathbf X}^{-1}
 =
 \begin{pmatrix}
 -i\sqrt{a-|z_1|^2}&z_1\\ \bar z_1&-i\sqrt{a-|z_1|^2}
 \end{pmatrix},
 \qquad
 \cG_{\mathbf Y}^{-1}
 =
 \begin{pmatrix}
 -i\sqrt{c-|z_2|^2}&z_2\\ \bar z_2&-i\sqrt{c-|z_2|^2}
 \end{pmatrix}.
\]
The $4\times4$ inversion \eqref{eq:G2_pair} can be carried out in closed
form and gives
\begin{equation}\label{eq:D_pair_Ginibre}
 D_{\mathbf A,\mathbf A+\mathbf B}(z_1,z_2)
 =
 \frac{a\,(c-|z_2|^2)-c\,\bar z_1(z_1-z_2)}
 {a\left[\,c\,|z_1-z_2|^2+b\,(c-|z_2|^2)\right]}.
\end{equation}
Only the common variance $a$ enters $\Rtwo_{\mathbf A,\mathbf A}$,
whereas $b$ enters through the Green function of $\mathbf A+\mathbf B$
and regularizes the denominator.  Two Wirtinger
derivatives and \eqref{eq:Q_conditional}, with
$\rho_{\mathbf A}=(\pi a)^{-1}$ and $\rho_{\mathbf A+\mathbf B}=(\pi c)^{-1}$
on the respective disks, yield
\begin{equation}\label{eq:Q_Ginibre}
 N\,\E\!\left[
 \mathcal Q_{ij}^{\mathbf A,\mathbf A+\mathbf B}
 \,\middle|\,
 \lambda_i=z_1,\ \mu_j=z_2
 \right]
 =
 -\,\frac{c^3(c-z_1\bar z_2)
 \left[c\,|z_1-z_2|^2+b\left(2z_1\bar z_2-|z_2|^2-c\right)\right]}
 {\left[c\,|z_1-z_2|^2+b\,(c-|z_2|^2)\right]^3} .
\end{equation}
For $b\downarrow0$ this reduces to
$-a\,(a-z_1\bar z_2)/|z_1-z_2|^4$,
the conditional Chalker--Mehlig overlap at variance $a$.  For $b>0$ the singularity at $z_1=z_2$ is removed: on the
diagonal
\begin{equation}\label{eq:Q_diagonal}
 N\,\E\!\left[
 \mathcal Q_{ij}^{\mathbf A,\mathbf A+\mathbf B}
 \,\middle|\,
 \lambda_i=z,\ \mu_j=z
 \right]
 =
 \frac{c^3}{b^2\,(c-|z|^2)},
\end{equation}
which is positive and of order $b^{-2}$ as $b\downarrow0$.  For $z_1=0$,
\begin{equation}\label{eq:Q_anchored}
 N\,\E\!\left[
 \mathcal Q_{ij}\mid\lambda_i=0,\ \mu_j=z_2
 \right]
 =
 \frac{c^4\left(bc-a|z_2|^2\right)}
 {\left[c\,|z_2|^2+b\,(c-|z_2|^2)\right]^3},
\end{equation}
which is real and changes sign at $|z_2|^2=bc/a$.

\begin{figure}[t]
 \centering
 \includegraphics[width=\textwidth]{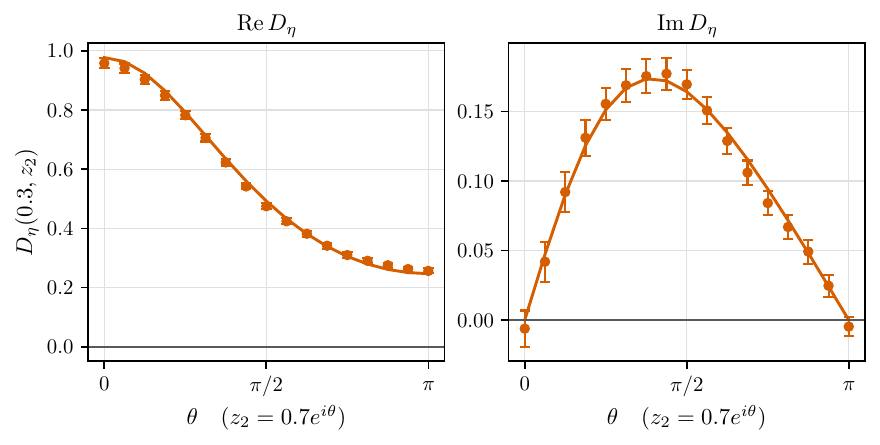}
 \caption{Real and imaginary parts of
 $D_\eta(z_1,z_2)
 =\frac1N\Tr\bigl[
 \mathbf G_{21}^{\mathbf A}(-i\eta,z_1)\,
 \mathbf G_{12}^{\mathbf A+\mathbf B}(-i\eta,z_2)
 \bigr]$
 against $\theta$, for independent complex Ginibre matrices
 $\mathbf A$ and $\mathbf B$,
 $z_1=0.3$, $z_2=0.7e^{i\theta}$, $a=1$, $b=0.5$ and $\eta=0.12$.
 Line: $(11),(22)$ entry of \eqref{eq:G2_pair} at $\omega=-i\eta$.
 Points: same trace averaged over $25$ pairs of size $N=60$.}
 \label{fig:mixed-correlator}
\end{figure}

Figure~\ref{fig:mixed-correlator} compares \eqref{eq:G2_pair} at
$\omega_1=\omega_2=-i\eta$ with simulations at $z_1=0.3$ and
$z_2=0.7e^{i\theta}$.  The $\eta\to0$ derivatives of this entry give
\eqref{eq:Q_Ginibre}.

\section{Singular-vector overlaps}\label{sec:singular-overlaps}

\subsection{Two matrices and two singular-vector bases}

Let $\mathbf A$ and $\mathbf A+\mathbf B$ be two non-Hermitian matrices and
write their singular-value decompositions as
\[
 \mathbf A=\sum_i s_i\,u_i v_i^*,
 \qquad
 \mathbf A+\mathbf B=\sum_j t_j\,\widetilde u_j\widetilde v_j^*.
\]
The conditional squared overlaps of their left and right singular vectors
are
\begin{align}
 \Phi^L_{\mathbf A,\mathbf A+\mathbf B}(s,t)
 &:=
 N\,\E\!\left[
 |\langle u_i,\widetilde u_j\rangle|^2
 \,\middle|\,s_i=s,\ t_j=t
 \right],
 \label{eq:Phi_cond_def}\\
 \Phi^R_{\mathbf A,\mathbf A+\mathbf B}(s,t)
 &:=
 N\,\E\!\left[
 |\langle v_i,\widetilde v_j\rangle|^2
 \,\middle|\,s_i=s,\ t_j=t
 \right].
\end{align}
The independent-basis value is one.  The $(11),(11)$ coefficient of
\eqref{eq:G2_pair} gives $\Phi^L$, while the $(22),(22)$ coefficient gives
$\Phi^R$; neither coefficient mixes left and right singular vectors.

\subsection{Resolvent representation}

For a matrix $\mathbf M$, define
\begin{equation}\label{eq:g1_singular}
 g_{\mathbf M}(\omega)
 :=
 \lim_{N\to\infty}\frac1N\Tr\!\left[
 \omega(\omega^2\1-\mathbf M\mathbf M^*)^{-1}\right]
 =
 \int_0^\infty
 \frac{\omega}{\omega^2-s^2}\rho_{\mathbf M}(s)\,\mathrm ds,
\end{equation}
where $\rho_{\mathbf M}$ is its limiting singular-value density.  Set
\begin{equation}\label{eq:psi_singular}
 \psi^L(\omega_1,\omega_2)
 =
 \frac1N\E\Tr\!\left[
 \omega_1(\omega_1^2\1-\mathbf A\mathbf A^*)^{-1}
 \omega_2(\omega_2^2\1-(\mathbf A+\mathbf B)(\mathbf A+\mathbf B)^*)^{-1}
 \right].
\end{equation}
Its singular-vector expansion is
\begin{equation}
 \psi^L(\omega_1,\omega_2)
 =
 \frac1N\E\sum_{i,j}
 \frac{\omega_1\omega_2|\langle u_i,\widetilde u_j\rangle|^2}
 {(\omega_1^2-s_i^2)(\omega_2^2-t_j^2)}.
\end{equation}
The analogous expression with $\mathbf M^*\mathbf M$ gives $\psi^R$.
For $\eta\downarrow0$ with $\eta\gg N^{-1}$, the
Sokhotski--Plemelj formula yields
\begin{equation}\label{eq:Phi_from_psi}
 \Phi^L_{\mathbf A,\mathbf A+\mathbf B}(s,t)
 =
 \frac{2}{
 \pi^2\rho_{\mathbf A}(s)\rho_{\mathbf A+\mathbf B}(t)
 }\,
 \Re\!\left[
 \psi^L(s-i\eta,t+i\eta)-\psi^L(s-i\eta,t-i\eta)
 \right].
\end{equation}
Indeed,
$\omega(\omega^2-s^2)^{-1}
=\frac12[(\omega-s)^{-1}+(\omega+s)^{-1}]$, so its jump across the
positive real axis is $i\pi\delta(\omega-s)$; taking the real part in
\eqref{eq:Phi_from_psi} supplies the remaining factor $\pi/2$ from the
first argument.  This is the same extraction as in the Hermitian
construction of
\cite{BunBouchaudPotters2018Overlaps,potters2020first}.

\subsection{Ginibre matrices}

Assume that $\mathbf A$ and $\mathbf B$ are independent complex Ginibre
matrices with
\[
 \E|A_{ij}|^2=\frac aN,\qquad
 \E|B_{ij}|^2=\frac bN,\qquad c=a+b.
\]
Their singular-value transforms and densities are
\begin{equation}
 g_v(\omega)=\frac{\omega-\sqrt{\omega^2-4v}}{2v},
 \qquad
 \rho_v(s)=\frac{\sqrt{4v-s^2}}{\pi v}\,
 \mathbf1_{0<s<2\sqrt v}.
\end{equation}
Only the common matrix $\mathbf A$ enters $\Rtwo_{\mathbf A,\mathbf A+\mathbf B}$:
\[
 \Rtwo_{\mathbf A,\mathbf A+\mathbf B}
 =a\,\Rtwo_{\mathrm{Gin}}.
\]
Writing $q=g_a(\omega_1)g_c(\omega_2)$, the $(11),(11)$ entry of
\eqref{eq:G2_pair} is
\begin{equation}\label{eq:DLL_Ginibre_correlated}
 \psi^L(\omega_1,\omega_2)=\frac{q}{1-a^2q^2}.
\end{equation}
Set
$s=2\sqrt a\cos\theta$, $t=2\sqrt c\cos\phi$, and $r=a/c$.  Applying
\eqref{eq:Phi_from_psi} gives
\begin{equation}\label{eq:conditional_LL_Ginibre}
 \Phi^L_{\mathbf A,\mathbf A+\mathbf B}(s,t)
 =
 \frac{(1-r)\,[S(\theta-\phi)-S(\theta+\phi)]}
 {2\sin\theta\sin\phi},
 \qquad
 S(x)=\frac{\cos x}{1-2r\cos(2x)+r^2}.
\end{equation}
By left--right symmetry of complex Ginibre matrices, the same formula holds
for $\Phi^R$.  For $r=0$ it reduces to one; when $b\downarrow0$ it
concentrates on $s=t$.  Thus $\mathbf B$ changes the overlap through
$g_c$ and $\rho_c$, although it does not contribute to
$\Rtwo_{\mathbf A,\mathbf A+\mathbf B}$.

We fix two singular values $t$ of $\mathbf A+\mathbf B$ and plot
\eqref{eq:conditional_LL_Ginibre} as a function of the singular value $s$
of $\mathbf A$.  Markers are simulation averages.

\begin{figure}[t]
 \centering
 \includegraphics{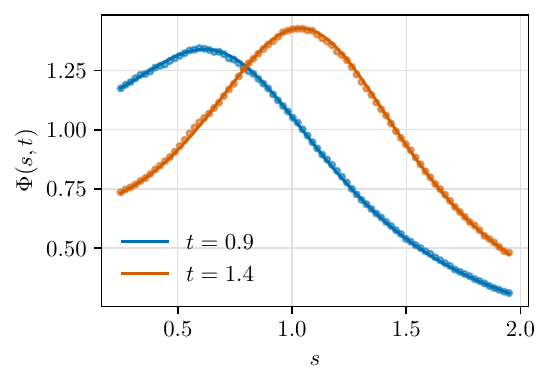}
 \caption{Left singular-vector overlaps between $\mathbf A$ and
 $\mathbf A+\mathbf B$ for $a=b=1$.  For fixed values of $t$ (legend),
 markers show simulations at $N=280$, using $220$ matrices, and solid lines show
 \eqref{eq:conditional_LL_Ginibre}.}
 \label{fig:singular-profile}
\end{figure}

\appendix
\section{Replica derivation of the two-point formula}\label{app:proof}

This appendix gives the replica derivation of
Result~\ref{result:master}.  We first work at sufficiently large
real $\omega_1$ and $\omega_2$, where the Gaussian integrals
converge.  We then calculate the terms quadratic in the
cross-Gram block around $C=0$ and continue the resulting formula
in $\omega_1$ and $\omega_2$.

\subsection{Gaussian representation of the hermitized resolvent}

Let $\mathbf C$ be a Hermitian positive-definite $2N\times2N$ matrix.  Then
\begin{equation}\label{eq:gaussian_det}
 \int e^{-\langle\varphi,\mathbf C\varphi\rangle}
 \frac{\mathrm d\varphi}{\pi^{2N}}
 =\frac1{\det\mathbf C},
 \qquad
 \int \varphi_p\bar\varphi_q\,
 e^{-\langle\varphi,\mathbf C\varphi\rangle}
 \frac{\mathrm d\varphi}{\pi^{2N}}
 =\frac{[\mathbf C^{-1}]_{pq}}{\det\mathbf C}.
\end{equation}
Introduce $n$ independent copies
$\varphi^{(1)},\dots,\varphi^{(n)}$.  For positive integer $n$, the
following integral is equal to
$[\mathbf C^{-1}]_{pq}(\det\mathbf C)^{-n}$.
The replica prescription consists in continuing this expression
to $n=0$:
\begin{equation}\label{eq:replica_identity}
 [\mathbf C^{-1}]_{pq}
 =
 \lim_{n\to0}
 \int
 \varphi_p^{(1)}\bar\varphi_q^{(1)}
 e^{-\sum_{k=1}^n
 \langle\varphi^{(k)},\mathbf C\varphi^{(k)}\rangle}
 \prod_{k=1}^n
 \frac{\mathrm d\varphi^{(k)}}{\pi^{2N}}.
\end{equation}
This is the usual replica prescription
\cite{mezard1987spin,potters2020first}.  We omit replica indices
when they are not needed and restore them in the Gram-matrix
calculation.

For the block matrix \eqref{eq:Cblock}, the Gaussian integral is
well defined for sufficiently large real $\omega$.

Split the field as $\varphi=(\varphi_1,\varphi_2)$, with
$\varphi_i\in\C^N$.  The quadratic form is
\begin{equation}\label{eq:quadratic_form}
 \langle\varphi,\mathbf C(\omega,z)\varphi\rangle
 =
 \omega\|\varphi_1\|^2+\omega\|\varphi_2\|^2
 +2\Re\langle\varphi_1,(z\1-\mathbf M)\varphi_2\rangle.
\end{equation}
The Gaussian identity therefore gives
\begin{equation}\label{eq:blocks_as_bilinears}
 \bigl[\mathbf C(\omega,z)^{-1}\bigr]_{\alpha\beta,\,ij}
 =
 \bigl\langle
 \varphi_{\alpha,i}\bar\varphi_{\beta,j}
 \bigr\rangle_{\mathbf C},
\end{equation}
where $\langle\cdot\rangle_{\mathbf C}$ denotes the normalized
Gaussian expectation with density proportional to
$e^{-\langle\varphi,\mathbf C\varphi\rangle}$.

\subsection{Two spectral points and the Gram matrix}

For each spectral point, introduce a separate set of Gaussian
integration variables.  We first write the contribution of one
replica at each point.  Let $\varphi_i^a$, where $a\in\{1,2\}$
labels $(\omega_a,z_a)$ and $i\in\{1,2\}$ labels the position
inside the hermitization, and set
\begin{equation*}
 \varphi_i^a=\sqrt N\,\psi_i^a.
\end{equation*}

The quadratic form becomes
\begin{equation}\label{eq:S0}
 S_0
 =
 N\sum_{a=1}^2
 \left[
 \sum_{i,j=1}^2(\mathbf Z_a)_{ij}\,
 \langle\psi_i^a,\psi_j^a\rangle
 -
 2\Re\langle\psi_1^a,\mathbf M\psi_2^a\rangle
 \right],
 \qquad
 \mathbf Z_a=\mathbf Z(\omega_a,z_a).
\end{equation}
For positive integer $n$, the same expression is summed over
the replicas.

The two off-diagonal terms in the first sum combine into
$2\Re(z_a\langle\psi_1^a,\psi_2^a\rangle)$, since
$\langle\psi_2^a,\psi_1^a\rangle
=\overline{\langle\psi_1^a,\psi_2^a\rangle}$.
For one pair of replicas, define
\begin{equation*}
 G^a_{ij}=\langle\psi_i^a,\psi_j^a\rangle,
 \qquad
 C_{ij}=\langle\psi_i^1,\psi_j^2\rangle.
\end{equation*}
These are the blocks of the Gram matrix \eqref{eq:full_gram},
\begin{equation*}
 \mathbb G
 =
 \begin{pmatrix}
 G^1&C\\
 C^*&G^2
 \end{pmatrix}.
\end{equation*}
For two independent Gaussian copies at fixed $\mathbf M$, the
Gaussian identity gives
\begin{equation}\label{eq:resolvents_as_C}
 \frac1N\sum_{i,j}
 [\mathbf C(\omega_1,z_1)^{-1}]_{\alpha\beta,ij}
 [\mathbf C(\omega_2,z_2)^{-1}]_{\gamma\delta,ji}
 =
 N\left\langle
 C_{\beta\gamma}\overline{C_{\alpha\delta}}
 \right\rangle.
\end{equation}
The replica prescription allows us to average this expression
over $\mathbf M$.  We therefore calculate the second moments of
$C$ from the terms quadratic in $C$ around $C=0$.

\subsection{Haar average}

Assume that $\mathbf M$ is rotationally invariant.  In
\eqref{eq:S0}, the matrix $\mathbf M$ appears only in the terms
$\langle\psi_1^a,\mathbf M\psi_2^a\rangle$.  By
Definition~\ref{def:H2}, the Haar average is
\begin{equation}\label{eq:haar_step}
 \E_{\mathbf U}
 \exp\left\{
 2N\Re\sum_{a=1}^2
 \langle\psi_1^a,
 \mathbf U\mathbf M\mathbf U^*\psi_2^a\rangle
 \right\}
 =
 \exp\bigl\{2N\Htwo_{\mathbf M}(\mathbb G)\bigr\}.
\end{equation}

After this average, the integrand depends on the fields only
through their Gram matrix.

\subsection{The block-diagonal solution}

Consider the solution $C=0$.  The two spectral points then
decouple, and each point satisfies the one-point stationarity
condition \eqref{eq:one_point_relation}.  Denote its solution by
$\cG_a$.  With our convention for the scalar product, the Gram
entries at this solution are
\begin{equation}\label{eq:gram_vs_green}
 G^a_{ij}=(\cG_a)_{ji}.
\end{equation}
Thus the arguments of the one-point transform are
$\alpha=\sqrt{G^a_{11}G^a_{22}}=\mathfrak g_1$ and
$\beta=G^a_{12}=(\cG_a)_{21}=\mathfrak g_2$.  In the initial
real $\omega_a$ domain, $(\cG_a)_{12}=\overline{\mathfrak g_2}$,
which fixes the placement of $\mathcal R_2=2\partial_\beta\Ha$
in $\mathbf R(\cG)$ as in \eqref{eq:matrix_ZGR}.

\subsection{Fluctuations of the cross-Gram block}

We now calculate the quadratic term in $C$.  This can be done
directly from the change of variables from the vectors to their
Gram matrix.

For positive integer $n$, collect the $4n$ vectors into an
$N\times4n$ matrix $\Psi$ and write
\begin{equation*}
 \mathbb G^{(n)}=\Psi^*\Psi.
\end{equation*}
For $4n\leq N$, and for an integrand depending on $\Psi$ only
through $\mathbb G^{(n)}$, the change of variables gives
\begin{equation}\label{eq:gram_jacobian}
 \int F(\Psi^*\Psi)\,\mathrm d\Psi
 =
 K_{N,n}
 \int_{\mathbb G^{(n)}>0}
 F(\mathbb G^{(n)})
 \bigl(\det\mathbb G^{(n)}\bigr)^{N-4n}
 \,\mathrm d\mathbb G^{(n)},
\end{equation}
where $K_{N,n}$ is independent of the Gram matrix.  The factor
introduced by the rescaling $\varphi=\sqrt N\,\psi$ is also
independent of the Gram matrix and is included in $K_{N,n}$.

Write the full replicated Gram matrix in block form,
\begin{equation*}
 \mathbb G^{(n)}
 =
 \begin{pmatrix}
 \mathbb G^{(n)}_1&\mathcal C\\
 \mathcal C^*&\mathbb G^{(n)}_2
 \end{pmatrix},
\end{equation*}
where the diagonal blocks have size $2n\times2n$.  The determinant
identity gives
\begin{equation*}
 \det\mathbb G^{(n)}
 =
 \det\mathbb G^{(n)}_1\,
 \det\mathbb G^{(n)}_2\,
 \det\left(
 I-
 (\mathbb G^{(n)}_2)^{-1}
 \mathcal C^*
 (\mathbb G^{(n)}_1)^{-1}
 \mathcal C
 \right).
\end{equation*}
Consequently, for fixed positive-definite diagonal blocks,
\begin{align}
 \log\det\mathbb G^{(n)}
 ={}
 &\log\det\mathbb G^{(n)}_1
 +
 \log\det\mathbb G^{(n)}_2
 \nonumber\\
 &-
 \operatorname{Tr}\left[
 (\mathbb G^{(n)}_1)^{-1}
 \mathcal C
 (\mathbb G^{(n)}_2)^{-1}
 \mathcal C^*
 \right]
 +
 O(\|\mathcal C\|^4).
 \label{eq:gram_expansion}
\end{align}

We take the solution for which scalar products between different
replicas vanish.  The diagonal blocks are then
\begin{equation*}
 \mathbb G^{(n)}_a
 =
 \operatorname{diag}(G^a,\ldots,G^a),
\end{equation*}
with $n$ copies of the one-point Gram matrix.  Write $C^{rs}$ for
the $2\times2$ block connecting replica $r$ at the first spectral
point to replica $s$ at the second.  The quadratic term from the
determinant is
\begin{equation*}
 -(N-4n)\sum_{r,s=1}^n
 \operatorname{Tr}\left[
 (G^1)^{-1}C^{rs}(G^2)^{-1}(C^{rs})^*
 \right].
\end{equation*}
For the pair of replicas used in the observable, we write
$C=C^{11}$.  In the replica limit, the coefficient becomes $N$.
The diagonal Gram blocks can be evaluated at their stationary
values because their fluctuations first couple to $C$ through
terms of order $\delta G\,C^2$.

Expanding this expression gives
\begin{equation*}
 \operatorname{Tr}\left[
 (G^1)^{-1}C(G^2)^{-1}C^*
 \right]
 =
 \sum_{a,b,c,d}
 \overline{C_{ab}}\,
 \bigl[(G^1)^{-1}\bigr]_{ac}
 \bigl[(G^2)^{-1}\bigr]_{db}
 C_{cd}.
\end{equation*}
The inverse of this quadratic form gives the covariance of the
corresponding Gaussian integral.  Using \eqref{eq:gram_vs_green},
we obtain
\begin{equation}\label{eq:decoupled_covariance}
 N\,\E_0\bigl[C_{ab}\overline{C_{cd}}\bigr]
 =
 (\cG_1)_{ca}(\cG_2)_{bd}.
\end{equation}
Here $\E_0$ denotes the Gaussian expectation obtained from the
quadratic term above.

Write
\begin{equation*}
 \cG_1
 =
 \begin{pmatrix}
 x_1&u_1\\
 v_1&x_1
 \end{pmatrix},
 \qquad
 \cG_2
 =
 \begin{pmatrix}
 x_2&u_2\\
 v_2&x_2
 \end{pmatrix}.
\end{equation*}
Assemble the column of Definition~\ref{def:R2} in the order
\begin{equation*}
 \bm c
 =
 \begin{pmatrix}
 C_{21}\\
 C_{22}\\
 C_{11}\\
 C_{12}
 \end{pmatrix}.
\end{equation*}
Equation~\eqref{eq:decoupled_covariance} then gives
\begin{equation*}
 N\,\E_0[\bm c\,\bm c^*]
 =
 \begin{pmatrix}
 x_1x_2&x_1u_2&u_1x_2&u_1u_2\\
 x_1v_2&x_1x_2&u_1v_2&u_1x_2\\
 v_1x_2&v_1u_2&x_1x_2&x_1u_2\\
 v_1v_2&v_1x_2&x_1v_2&x_1x_2
 \end{pmatrix}
 =
 \cG_1\otimes\cG_2.
\end{equation*}
The last equality uses the ordering above and the equal diagonal
entries of each $\cG_a$.

Thus the quadratic term supplied by the Gram-matrix measure has
covariance
\begin{equation}\label{eq:Sigma}
 \E_0\bigl[\bm c\,\bm c^*\bigr]
 =
 \frac1N(\cG_1\otimes\cG_2).
\end{equation}
In particular, $C$ fluctuates on the scale $N^{-1/2}$ at the
block-diagonal solution.

\subsection{Quadratic correction and two-point covariance}

The remaining contribution comes from the Haar average.
Using \eqref{eq:Fexpansion}, its expansion around $C=0$ is
\begin{equation}\label{eq:interaction}
 2N\Htwo_{\mathbf M}(\mathbb G)
 =
 2N\Ha_{\mathbf M}(G^1)
 +
 2N\Ha_{\mathbf M}(G^2)
 +
 N\bm c^*
 \Rtwo_{\mathbf M}(A,B)
 \bm c
 +
 o(N\|C\|^2).
\end{equation}
For the block $C^{11}$, we use the quadratic coefficient of
\eqref{eq:Fexpansion}, evaluated at the one-point solutions.
At $C=0$, the
arguments are $A=\cG_1$ and $B=\cG_2$, with the convention of
Definition~\ref{def:R2}.  The first two terms are already
included in the one-point equations.  Since $C$ is of order
$N^{-1/2}$, the quadratic term is of order one; higher-order
terms are neglected.  Combining this term with the determinant
contribution gives
\begin{equation*}
 -N\bm c^*
 \left[
 (\cG_1\otimes\cG_2)^{-1}
 -
 \Rtwo_{\mathbf M}(\cG_1,\cG_2)
 \right]
 \bm c.
\end{equation*}
For values of the parameters where the matrix in brackets is
positive definite, the Gaussian integral gives, at leading order,
\begin{equation}\label{eq:covariance_result}
 N\,\E_{\mathrm{quad}}\bigl[\bm c\,\bm c^*\bigr]
 =
 \left[
 (\cG_1\otimes\cG_2)^{-1}
 -
 \Rtwo_{\mathbf M}(\cG_1,\cG_2)
 \right]^{-1}
 =
 \Gtwo_{\mathbf M}.
\end{equation}
This is \eqref{eq:G2_master}.

The resulting formula is then analytically continued in
$\omega_1$ and $\omega_2$ wherever the continuation exists.

The two-matrix formula \eqref{eq:G2_pair} follows from the same
calculation, with $\mathbf M$ replaced by $\mathbf X$ in the first
insertion and by $\mathbf Y$ in the second.  The Haar contribution
is given by \eqref{eq:H2_pair}, while the determinant calculation
is unchanged, with
\begin{equation*}
 \cG_1=\cG_{\mathbf X}(\omega_1,z_1),
 \qquad
 \cG_2=\cG_{\mathbf Y}(\omega_2,z_2).
\end{equation*}

\section*{Acknowledgements}
This research was conducted within the Econophysics
\& Complex Systems Research Chair, under the aegis of
the Fondation du Risque, the Fondation de l'\'Ecole
polytechnique, the \'Ecole polytechnique, and Capital Fund
Management.
Generative AI tools were used during the preparation of this
manuscript solely to support language editing, reformulation, and
clarity.

\bibliographystyle{plain}
\bibliography{article_references}

\end{document}